\documentclass[12pt,preprint]{aastex}

\usepackage{amsmath,psfig,graphicx}

\def\rmd{{\rm d}}
\def\lsim{\mathrel{\rlap{\lower4pt\hbox{\hskip1pt$\sim$}}
\raise1pt\hbox{$<$}}}
\def\gsim{\mathrel{\rlap{\lower4pt\hbox{\hskip1pt$\sim$}}
\raise1pt\hbox{$>$}}}

\def\draftmark {
}
\newcommand{\erf}[1]{(\ref{#1})} \newcommand{\fl}{\hspace*{-6pc}}

\begin{document}

\draftmark

\title{Gravitational radiation timescales for extreme mass ratio 
inspirals}

\author{Jonathan R.\ Gair} \affil{Institute of Astronomy, Madingley Road, Cambridge, CB3 0HA, UK}

\author{Daniel J.\ Kennefick} \affil{Theoretical
Astrophysics, California Institute of Technology, Pasadena, CA 91125, USA
and Department of Physics, University of Arkansas, Fayetteville, AR
72701, USA}

\author{Shane L. Larson}\affil{Center for Gravitational Wave Physics,
The Pennsylvania State University, University Park, PA 16802, USA}


\begin{abstract}
The capture and inspiral of compact stellar objects into massive black
holes is an important source of low-frequency gravitational waves (with frequencies $\sim 1\--100$mHz), such as those that might be detected by the planned Laser Interferometer Space Antenna (LISA).
Simulations of stellar clusters designed to study this problem
typically rely on simple treatments of the black hole encounter which
neglect some important features of orbits around black holes, such as
the minimum radii of stable, non-plunging orbits.  Incorporating an
accurate representation of the orbital dynamics near a black hole has
been avoided due to the large computational overhead.  This paper
provides new, more accurate, expressions for the energy and angular
momentum lost by a compact object during a parabolic encounter with a
non-spinning black hole, and the subsequent inspiral lifetime.  These results
improve on the Keplerian expressions which are now commonly used and
will allow efficient computational simulations to be performed that
account for the relativistic nature of the spacetime around the central black hole in the system.
\end{abstract}

\keywords{Stellar captures -- loss cone -- gravitational radiation -- inspirals}

\section{Introduction}\label{sec:Intro}

The event rate for detection of extreme mass ratio inspirals (EMRIs)
with LISA depends on the efficiency of capture of compact objects by massive black holes in galactic nuclei.  Current event rate estimates \cite{emri04} are derived using results from stellar dynamics simulations of nuclear star clusters
\cite{hils95,SigRees97,Freitag01,Freitag02,Ivanov02,Freitag03,hopman05}, in which the clusters are dynamically evolved over the lifetime of a galaxy.  Of central importance to these simulations is the estimate of the effect of gravitational radiation on the evolution of a particle's orbital parameters as a function of
proximity to the central black hole.  The comparison of gravitational
radiation timescales against cluster interaction timescales is a way
to quantify whether a star stays bound to the black
hole (eventually spiralling in to its death) or if it returns to the
parent star cluster and is scattered away from the black hole by
encounters with other stars. Scattering by other stars can also put the star onto a quasi-radial orbit so that it plunges into the black hole directly rather than inspiralling gradually. LISA will only be able to detect long lived inspirals \cite{emri04}, so it is important to quantify the fraction of captures that terminate in these two distinct ways. Simulations suggest that a significant fraction of captures will end in quasi-radial plunges rather than inspirals \cite{hils95,hopman05}, which will impact the LISA event rate.

For any given encounter with the central black hole, the gravitational
radiation inspiral lifetime $\tau_{gw}$ of a member of the population
is compared against the two body relaxation timescale $\tau_{rlx}$ with
other particles in the simulation.  For an orbit of eccentricity $e$, if $\tau_{gw} < (1-e) \, \tau_{rlx}$, the particle is removed from the simulation, and its orbital parameters are used to estimate the strength of the gravitational waves which might reach our detectors.  If the relaxation timescale is such that $(1-e)\,\tau_{rlx} < \tau_{gw}$, then two body encounters with other cluster members will alter the pericenter distance of a particle's orbit before gravitational radiation reaction causes it to merge with the
central black hole \cite{SigRees97, Ivanov02}.


The conventional method for treating gravitational radiation in these
simulations is to use the formalism of Peters and Mathews \cite{PM63}
and Peters \cite{peters64}, which assumes the particles and central
masses are point-like, Newtonian objects and that the particle orbits
are Keplerian trajectories.  This framework does not account for the
fact that the central body is a black hole, and that the orbits and
orbital evolution can be decidedly non-Keplerian in nature.  This
paper presents an improvement to the traditional Peters and Mathews treatment, based on perturbative calculations.  By exploiting the extreme mass ratio of
the system, the inspiralling object can be regarded as a small
perturbation of the spacetime of the central black hole.  While black
hole perturbation theory is well understood \cite{poisson04}, the solution requires extensive numerical calculations for general orbits.  However, the orbits of interest in stellar dynamics calculations are usually
highly eccentric, and for such orbits the inspiral timescale can be estimated to an accuracy of $\sim1\%
$ (depending on the precise eccentricity at capture) simply by knowing the change in eccentricity during the first encounter with the black hole.

Data for the energy and angular momentum change on a parabolic orbit
are available in the literature \cite{martel04}, and can be used as a
starting point for treating the dynamics of extreme mass ratio
inspirals.  Based on an understanding of the properties of geodesics
in the Schwarzschild spacetime, it is possible to fit a simple
function to this data; the corresponding formulae are all that is
required by stellar dynamics codes and represent a significant
improvement over the standard Keplerian treatment \cite{PM63}.

This paper is organised as follows.  Section \ref{sec:Fluxes} describes geodesics in the Schwarzschild spacetime
and provides fits for the total radiated energy $\Delta E$ and angular momentum $\Delta L$ on single parabolic encounters with the black hole. Section \ref{sec:Timescales} presents an expression for the inspiral lifetime based on these fits. In Section \ref{genorbits} we briefly discuss how the gravitational radiation losses from orbits of arbitrary (low) eccentricity can be estimated and finally, Section \ref{sec:Discussion} summarises the implications and possible applications of this work.

\section{Geodesics and gravitational wave fluxes for parabolic orbits}\label{sec:Fluxes}
The results of Peters and Mathews \cite{PM63} for the radiated energy
and angular momentum and the inspiral lifetime are simple to
implement, with a low associated computational cost.  This makes them
ideal for use in large simulations.  The formalism has the
disadvantage that it treats the binary components as point masses on
Keplerian orbits.  Captured stars that evolve into an extreme
mass ratio inspiral (EMRI) generally must pass very close to the black
hole, where the orbit is very non-Keplerian.  The Peters and Mathews
treatment neglects important features of the gravitational wave
emission due to the presence of the black hole.

One simple improvement that can be made is to use a ``semi-relativistic'' approximation, i.e., using the fully relativistic orbit in place of the Keplerian orbit, while
using an approximation for the corresponding gravitational wave
emission. This approach was first suggested by \cite{RufSas} and is explored extensively in a companion paper \cite{GKL2}. To compute things correctly, one must use black hole
perturbation theory and solve the Teukolsky equation \cite{poisson04}. The problem of radiation from orbits in the Schwarzschild spacetime was examined by \cite{CKP}. They provided useful asymptotic results for nearly circular and nearly plunging orbits and tabulated fluxes for orbits with a variety of periapses and eccentricities. However, their code worked in the frequency domain, which is not well suited for dealing with the highly eccentric orbits of interest in the capture problem. More recently, results for the fluxes of radiation from parabolic orbits in Schwarzschild have become available \cite{martel04}, which were computed using a time domain code. Using insight gained from studying the geodesic equations \cite{GKL2}, it is possible to derive a simple fitting
function for the energy and angular momentum emitted that matches the
perturbative results to within a fraction of a percent.  This function may be
implemented in stellar cluster simulations as easily as the Keplerian
expressions and for little additional computational cost.

The Schwarzschild geodesic equations for an equatorial orbit (without
loss of generality) are
\begin{eqnarray}
   \left(\frac{\rmd r}{\rmd\tau}\right)^2 & = & \left( \frac{E^{2}}{c^2} - c^2\right)
   +\frac{2\,GM}{r}\,\left(1+\frac{L_{z}^{2}}{c^2\,r^{2}}\right)
   -\frac{L_{z}^{2}}{r^{2}} \,, \label{rdot} \\
   \left(\frac{\rmd\phi}{\rmd\tau} \right) &=& \frac{L_{z}}{r^{2}} \,,
   \label{phidot} \\
   \left(\frac{\rmd t}{\rmd\tau}
   \right) &=& \frac{E}{\left(c^2-2\,GM/r\right)},  \label{tdot}
\end{eqnarray}
where $\tau$ is the proper time along the geodesic, $L_{z}$ is the
conserved specific angular momentum of the particle\footnote{For equatorial orbits in Schwarzschild, the z-component of the angular momentum, $L_{z} = L$, the total angular momentum. We maintain the notation $L_{z}$ in order to facilitate future comparisons with orbits in Kerr spacetimes, for which $L_z$ is conserved, but not $L$.}, $E$ is the conserved specific energy and $M$ is the mass of the central black hole. In the weak field ($r \gg GM/c^2$ or $L_{z} \gg GM/c$), these reduce to the usual Keplerian equations of motion, the geodesic is a conic section and there is a well defined periapse and eccentricity. These are related to the orbital energy and angular momentum by
\begin{eqnarray}
\frac{E}{c^2} &=& \sqrt{1-\frac{GM\,(1-e^K)}{c^2\,r_{p}^K}}\,, \label{EofrpeKep} \\  L_{z} &=& \sqrt{(1+e^K)\,GM\,r_{p}^K}\,. \label{LofrpeKep}
\end{eqnarray}
We have used ``K'' to denote a parameter defined in the Keplerian way. In the strong field, relativistic effects cause the orbit to deviate from Keplerian motion and the normal notion of eccentricity, as a geometrical parameter
characterizing the shape of a conic section, is not valid. However, we can still define a relativistic orbital periapse, $r_p^R$, as the Schwarzschild radial coordinate of the inner turning point of the motion, and we can define a relativistic eccentricity, $e^R$, from $r_p^R$ and the Schwarzschild coordinate of the outer turning point of the motion (the relativistic apapse, $r_a^R$), using the usual equation
\begin{equation}
e^R = \frac{r_{a}^R - r_{p}^R}{r_{a}^R + r_{p}^R}.
\end{equation}
The relationship between the relativistic parameters and the orbital energy and angular momentum is
\begin{eqnarray}
    \frac{E}{c^2} &=& \sqrt{1-\frac{GM\,(1-e^R)((1+e^R)\,c^2\,r_{p}^R-4\,GM)}{c^2\,r_{p}^R((1+e^R)\,
    c^2\,r_{p}^R-(3+{e^R}^{2})\,GM)}}\,, \label{EofrpeRel} \\
   L_{z} &=& \frac{(1+e^R)\,c\,\sqrt{GM}\,r_{p}^R}{\sqrt{(1+e^R)\,c^2\,
   r_{p}^R-(3+{e^R}^{2})\,GM}} \,.
   \label{LofrpeRel}
\end{eqnarray}
It is important to note the differences between \erf{EofrpeRel}--\erf{LofrpeRel} and \erf{EofrpeKep}--\erf{LofrpeKep}. In simulations of stellar clusters the parameters of orbits that pass close to the black hole are generally computed using the Keplerian relations. However, this is not a good approximation for orbits that pass within a few Schwarzschild radii of the black hole. The energy and angular momentum are well defined out in the cluster where the orbital parameters are determined. Equating the right hand side of \erf{EofrpeRel} with that of \erf{EofrpeKep}, and similarly for the right hand sides of \erf{LofrpeRel} and \erf{LofrpeKep}, we can deduce a relationship between the Keplerian eccentricity and periapse and the relativistic eccentricity and periapse, for orbits which have a specified energy and angular momentum. We are mainly interested in highly eccentric orbits, so we work to linear order in $(1-e^{K})$.
\begin{eqnarray}
r_p^R &=& \frac{r_p^K}{2}\left(1+\sqrt{1-\frac{8GM}{c^2\,r_p^K}}\right) \nonumber \\
&& - \frac{(1-e^K)}{2}\,\frac{GM}{c^2}\,\frac{\sqrt{r_p^K(r_p^K-8GM/c^2)}+r_p^K-8GM/c^2}{r_p^K-8GM/c^2} \nonumber \\
1-e^R &=& \frac{(1-e^K)}{2}\left(1+\sqrt{1-\frac{8GM}{c^2\,r_p^K}}\right)
\label{KepRelTransform}
\end{eqnarray}
The energy and angular momentum losses given below are expressed in terms 
of the relativistic (``R'') parameters, so it is important to use equation \erf{KepRelTransform} to convert Newtonian parameters into their relativistic counterparts when evaluating gravitational wave fluxes. Indeed, even when using Peters and Mathews level approximations one should use the relativistic rather than the Keplerian parameters for better results. This approach was used in \cite{hopman05} and we discuss it in more detail in Section~\ref{genorbits}.

For the rest of this section we will concentrate on parabolic orbits, i.e., orbits for which $e^K = e^R = 1$. A parabolic orbit has $E=c^2$ and is uniquely parameterised by its periapse (or angular momentum). The angular momentum is related to the periapse by
\begin{equation}
   L_{z} = \frac{\sqrt{2\,GM}\,c\,r_{p}^R}{\sqrt{c^2\,r_{p}^R-2\,GM}} .
   \label{Lofrp}
\end{equation}
This should be compared to the Keplerian relation, $L_{z}=\sqrt{2GM\,r_p^K}$. The radial geodesic equation for a parabolic orbit is
\begin{equation}
   \left(\frac{\rmd r}{\rmd\tau}\right)^2 =
   \frac{2\,GM}{r^{3}}
\,\left(r-r_{p}^R\right)\left(r-\frac{2\,G\,M\,r_{p}^R} {c^2\,r_{p}^R-2\,G\,M}\right)
\label{rdotPar}
\end{equation}
For any given eccentricity, there is a minimum value for the periapse
below which the orbit plunges directly into the black hole.  This
occurs when the two inner turning points of the geodesic equation
coincide.  For parabolic orbits this is $r_{p}^R=4\,GM/c^2$.  A
geodesic with precisely this periapse asymptotically approaches a
circular orbit as it nears the periapse, and spends an infinite amount
of time whirling around the black hole.  The asymptotic circular orbit
is an unstable orbit of the gravitational potential.  Cutler,
Kennefick and Poisson \cite{CKP} call this orbit the `separatrix',
since it separates bound from plunging orbits in phase space.

In calculating the energy and angular momentum lost from an orbit, we must make the assumption of
adiabaticity, i.e., that the timescale over which the parameters of the orbit change
significantly due to gravitational radiation is much longer than the timescale
of the orbit. This is valid in the extreme mass ratio limit, $m/M \ll 1$. Under
this approximation, we treat the orbit as an exact geodesic of the spacetime,
compute the corresponding radiation fluxes and then update the orbital
parameters to a new geodesic before repeating this procedure. A particle on a
separatrix orbit would radiate an infinite amount of gravitational radiation, as
it spends an infinite time ``whirling'' around the black hole on a nearly
circular orbit. In this case, adiabaticity breaks down and it is wrong to
neglect the effect of radiation reaction. In practice a particle that starts on such an orbit would plunge into the black hole in a finite time. However, one still expects the energy and angular momentum losses to diverge as the separatrix is approached. 

\begin{figure}
\centerline{\includegraphics[keepaspectratio=true,height=3.in,
                             angle=0]{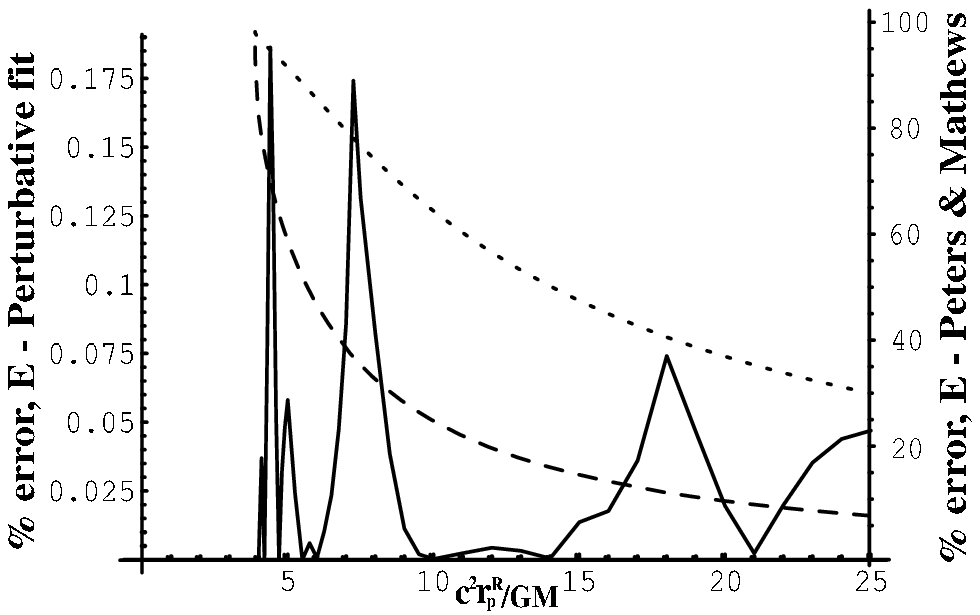}}

\vspace{0.2in}
\centerline{\includegraphics[keepaspectratio=true,height=3.in,
                             angle=0]{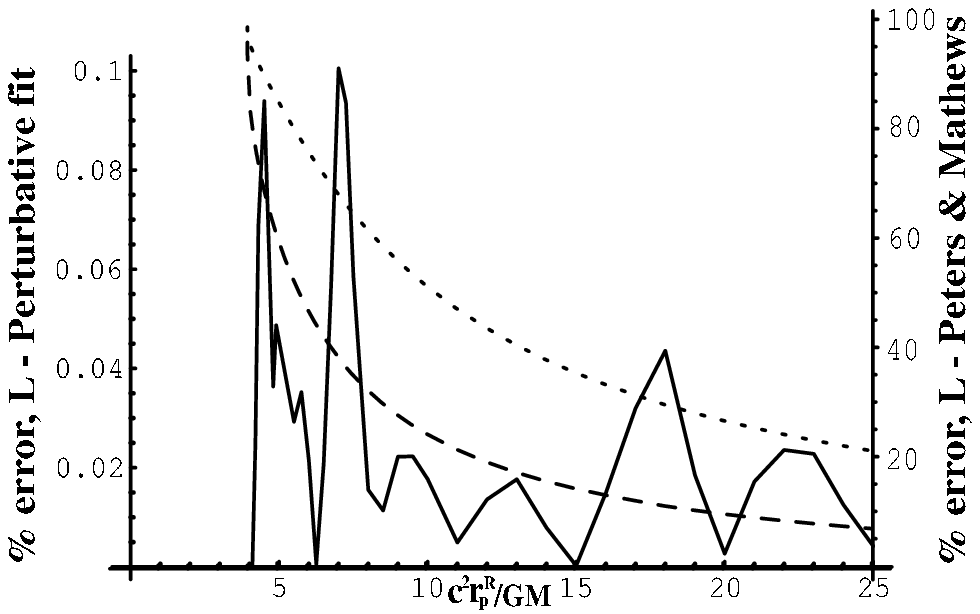}}

\caption{Accuracy of fit to relativistic fluxes -- this figure shows
the absolute percentage error when using the fitting function
described in the text (solid line) to approximate the energy (upper) and angular
momentum (lower) fluxes tabulated in Martel 2004. For comparison, we also show the error from using the Peters and Mathews result \erf{PMpardEdL}, evaluated for Keplerian (``K'') parameters (dotted line) and for relativistic (``R'') parameters (dashed line). Two different scales have been used -- the left hand scale applies to the errors in the fitting function, while the right hand scale applies to the errors in both applications of Peters and Mathews. The horizontal axis is the value of the relativistic periapse, $c^2\,r_p^R/GM$, of the geodesic in question.}
\label{martelfitfig}
\end{figure}

During the whirl phase, the orbit is almost circular, and so the total energy and angular momentum radiated will be approximately proportional to the number of ``whirls'' (i.e., complete revolutions in $\phi$) that the orbit undergoes. Counting the number of whirls indicates that, for a parabolic orbit, the total radiated energy and angular momentum will diverge like the logarithm of $r_p^R - 4GM/c^2$ near the separatrix. The derivation of this result is given in more detail in \cite{GKL2}, and was also discussed in \cite{CKP}. For orbits that do not come near the black hole, the Keplerian approximation is expected to be valid. Therefore, in the limit $r_p^R \rightarrow \infty$, the energy and angular momentum radiated, $\Delta E$ and $\Delta L_{z}$, should approach the Peters and Mathews results:
\begin{equation}
\Delta E = -\frac{85\,\pi}{12\,\sqrt{2}} \,c^2\, \frac{m}{M} \,
   \left(\frac{c^2\,r_{p}^R}{G\,M}\right)^{-\frac{7}{2}}, \qquad \Delta L_{z} =
-6\,\pi\,\frac{G\,m}{c}\,\left(\frac{c^2\,r_{p}^R}{G\,M}\right)^{-2} \ .
\label{PMpardEdL} 
\end{equation}
Our aim is to write a single expression for $\Delta E$ that can be used for any choice of $r_p^R$. Using the preceding arguments, we deduce that any such expression must diverge logarthmically near the separatrix at $r_p^R = 4GM/c^2$, and must recover \erf{PMpardEdL} in the limit $r_p^R \rightarrow \infty$. Denoting $y=c^2\,r_{p}^R/(GM)$, a functional form that has the correct behaviour in these two limits is
\begin{eqnarray}
  \nonumber \fl \frac{M}{m}\,\Delta X &=& F_{X}(y) \\ \nonumber &=& \left(\sum_{n=0}^{N}
  A^{X}_{n}\,\left(\frac{(y-4)}{y^{2}}\right)^{n}
  \right)\,\cosh^{-1}\left[1+B^{X}_{0} \,
  \left(\frac{4}{y}\right)^{N_{X}-1}\,
  \frac{1}{y-4}\right] \\ & & +
  \frac{(y-4)}{y^{1+
  \frac{N_{X}}{2}}} \,\sum_{n=0}^{N}C^{X}_{n}\,\left(\frac{
  (y-4)}{y^{2}}\right)^{n} +
  \frac{(y-4)}{y^{2+\frac{N_{X}}{2}}}\, \sum_{n=0}^{N-1}
  B^{X}_{n+1}\,\left(\frac{(y-4)}{y^{2}}\right)^{n}
\label{genfitform}
\end{eqnarray}
In this, $X$ is either $E/c^2$ or $c\,L_{z}/(G\,M)$ and we fix $N_{E}=7$, $N_{L_{z}}=4$ to give the correct leading order behaviour \erf{PMpardEdL} as $r_{p}^R\rightarrow\infty$. The parameter $N$ gives the order of the fit, i.e., the number of terms in the expansion that we use. To ensure that the fitting function asymptotically approaches \erf{PMpardEdL}, we impose a constraint on the coefficient $C_{0}^{X}$
\begin{equation}
C_{0}^{E} = -\frac{85\pi}{12\sqrt{2}}-64\,A_{0}^{E}\,\sqrt{2\,B_{0}^{E}}, \qquad C_{0}^{L_z} = -6\pi-8\,A_{0}^{L_z}\,\sqrt{2\,B_{0}^{L_z}}
\end{equation}
Further discussion of this fitting function is given in \cite{GKL2}. In that paper, we derive the fitting function coefficients that match the results of a ``semi-relativistic'' calculation. However, the most accurate calculation of energy and angular momentum fluxes requires solution of the Teukolsky equation. Data from such calculations is available in the literature for parabolic orbits around Schwarzschild black holes \cite{martel04}. Using the data from that paper, we were able to derive fitting function coefficients to use in \erf{genfitform} that recover the Teukolsky results extremely well. In fact, taking $N=2$ is sufficient for better than $0.2\% 
$ accuracy, and the corresponding fit coefficients are
\begin{eqnarray}
\nonumber A^{E}_{0}&=&-0.318434, \qquad A^{E}_{1}=-5.08198, \qquad
A^{E}_{2}=-185.48, \qquad B_{0}^{E} =0.458227, \\ \nonumber B^{E}_{1}&=&1645.79, \qquad B^{E}_{2}=8755.59, \qquad C^{E}_{0}=3.77465, \qquad C^{E}_{1}=-1293.27, \\ \nonumber C^{E}_{2}&=&-2453.55, \qquad  A^{L_{z}}_{0}=-2.53212, \qquad A^{L_{z}}_{1}=-37.6027, \qquad A^{L_{z}}_{2}=-1268.49, \\ \nonumber
B_{0}^{L_{z}}&=&0.671436, \qquad B^{L_{z}}_{1}=1755.51, 
B^{L_{z}}_{2}=9349.29, \qquad C^{L_{z}}_{0}=4.62465, \\   
C^{L_{z}}_{1}&=&-1351.44, \qquad C^{L_{z}}_{2}=-2899.02	
\label{martelfitcoeffs} 					
\end{eqnarray}
Figure~\ref{martelfitfig} shows the percentage error in using this
approximation over the range of periapse given by Martel. For comparison, we also show the error in using the Peters and Mathews result \erf{PMpardEdL}, evaluated for Keplerian and relativistic parameters. The error from using the fitting function \erf{genfitform} is significantly smaller than the difference between these fluxes and the Peters and Mathews results. The fluctuations in the error are due to the difference between a smooth function and noisy numerical results. The magnitude of the difference is everywhere smaller than the numerical error that Martel quotes (1\%
) and is a factor of approximately $1000$ smaller than the error using Peters and Mathews.

Expression \erf{genfitform} strictly applies only to parabolic orbits, i.e., with $e^R =1$. In realistic situations, the eccentricity at capture will be very high, but less than unity, $0<1-e_0^R << 1$. For such orbits, \erf{genfitform} can still be used and gives reliable results. The functional form fails if $r_p^R < 4\,GM/c^2$, but the last stable orbit (LSO) is related to the orbital eccentricity by $c^2\,{r_p^R}_{LSO} = 2GM(3+e^R)/(1+e^R)$ and so ${r_p^R}_{LSO} > 4\,GM/c^2$ for all $e^R < 1$. In fact, for non-parabolic orbits, a slightly better expression for $\Delta E$ is obtained by using \erf{genfitform} with $(y-4)$ replaced by $(y-2(3+e^R)/(1+e^R))$ and $(4/y)$ replaced by $2(3+e^R)/((1+e^R)\,y)$ (there is some discussion of suitable fitting functions for generic orbits in \cite{GKL2}). However, for extremely eccentric orbits, this change only makes a difference for orbits that are extremely close to the LSO.

\section{Inspiral Timescales}\label{sec:Timescales}
In stellar dynamics calculations which attempt to estimate the LISA
EMRI event rate, one must determine when a given particle is
captured by the central black hole.  Broadly speaking, a particle is captured when $\tau_{gw} \lsim (1-e)\,\tau_{rlx}$ (as before, $\tau_{gw}$ and $\tau_{rlx}$ are the timescales for gravitational wave inspiral and two body relaxation respectively). Heuristically, the picture is that if the orbital parameters evolve rapidly enough due to the emission of gravitational radiation, the star will spiral into the black hole (be ``captured'') before the cumulative perturbations to its orbit due to two-body encounters with other members of the cluster become large enough to put the star onto a new orbit which either does not come near the central black hole or plunges directly.

The canonical estimate of $\tau_{gw}$ is given by Peters
\cite{peters64} for a star which initially has semi-major axis $a_{0}$
and eccentricity $e_{0}$:
\begin{equation}
	\tau_{gw} = -\int_{0}^{e_{0}^K} \, \frac{1}{\rmd e/\rmd t} \, \rmd e =
\frac{12 c_{0}^{4}}{19 \beta}
	\int_{0}^{e_{0}^K} de \frac{e^{29/19} \left[ 1 +
	(121/304)e^{2}\right]^{1181/2299}}{\left(1 - e^{2}\right)^{3/2}}\ ,
	\label{petersTgw}
\end{equation}
where the constants $c_{0}$ and $\beta$ are given by
\begin{equation}
	c_{0} = \frac{(r_p^K)_{0} (1 + e_{0}^K)}{{e_{0}^K}^{12/19} \left[1 +
	(121/304){e_{0}^K}^{2} \right]^{870/2299}}\ , \qquad \beta =
\frac{G^{3}}{c^{5}}\frac{64}{5}M^2\,m\,.
	\label{petersconsts}
\end{equation}
In writing this and subsequent expressions in this section, we have 
assumed an extreme mass ratio, $M \gg m$ to set $M + m \approx M$.

Expression (\ref{petersTgw}) is derived by integration of the Peters
and Mathews fluxes over an inspiral. However, stars that become EMRI
events for LISA are captured with very high eccentricity (typically
$e\sim 0.9999$ or higher). In the limit $e_{0}^K\rightarrow 1$, expression (\ref{petersTgw}) becomes
\cite{peters64}
\begin{equation}
    \tau_{gw}(r_{p}^K,e_{0}^K) \approx \frac{24\,\sqrt{2}}{85}\,
	\frac{c^5}{G^{3}\,M^2\,m}\,\frac{{r_{p}^K}^{4}}{\sqrt{1-e_{0}^K}} .
   \label{tpetersAsym}
\end{equation}
This form of the timescale expression is used directly in some stellar
cluster simulations \cite{hopman05}, and is a very accurate approximation to the true timescale for inspirals that are initially highly eccentric. The divergence of the inspiral timescale as $e_{0}^K \rightarrow 1$ arises from the divergence of the orbital timescale (at fixed periapse) in the same limit
\begin{equation}
   T_{orb}(r_p,e_{0}) \approx \frac{2\,\pi}{\sqrt{G\,M}}\,
   \left(\frac{r_{p}}{1-e_{0}}\right)^{\frac{3}{2}}
\label{asymperiod}.
\end{equation}
In equation \erf{asymperiod} (and equations \erf{genedot} and \erf{genAsym} below), $r_p$ and $e$ can be either the relativistic or the Keplerian values, so superscripts have been omitted. Equation (\ref{asymperiod}) gives the dominant piece of the timescale even accounting for the presence of the black hole. For nearly parabolic orbits
\begin{equation}
   \frac{\rmd e}{\rmd t} \approx \frac{\Delta e(r_{p},e=1)}{T_{orb}(r_{p},e)} \approx
   \frac{\sqrt{GM}}{2\,\pi}\,\left(\frac{1-e}{r_{p}}
   \right)^{\frac{3}{2}}\,\Delta e(r_{p},e=1)
   \label{genedot}
\end{equation}
in which $\Delta e$ is the change in eccentricity on a single orbit (the initial pass). The timescale for inspiral is dominated by
\begin{equation}
   \tau_{gw}(r_{p},e_{0}) \approx -\frac{2}{\Delta
   e(r_{p}, e=1)}\,
   \frac{2\,\pi}{\sqrt{G\,M}}\,\frac{r_{p}^{\frac{3}{2}}}
   {\sqrt{1-e_{0}}}
   \label{genAsym}
\end{equation}
This expression is also given in \cite{hopman05}. In the Peters and Mathews approximation \cite{PM63}, $\Delta e^K(r_p^K,1) = -85\pi\,(G\,M)^{3/2} G\,m/(6\sqrt{2}\,c^5\,{r_p^K}^{5/2})$,
giving the result \erf{tpetersAsym}.  Using the fits to the
perturbative results, $\Delta e^R(r_{p}^R,e^R=1)$ may be expressed
\begin{eqnarray}
   \Delta e^R(r_{p}^R,e^R=1) & = & \frac{\partial e^R}{\partial E} \Delta
    E(r_{p}^R,e^R=1) + \frac{\partial e^R}{\partial L_{z}} \Delta
    L_{z}(r_{p}^R,e^R=1) \nonumber \\
    & = & 2\,\frac{r_{p}^R}{G\,M}\,\Delta E(r_{p}^R,e^R=1) \nonumber \\
    & = & 2\,\frac{c^2\,r_{p}^R}{G\,M}\,\frac{m}{M}\,F_{E}\left(\frac{c^{2}\,r_{p}^R}{GM}\right)
    \label{Deltae}
\end{eqnarray}
where the function $F_{E}(y)$ is the fit to the energy loss derived
earlier (\ref{genfitform}\--\ref{martelfitcoeffs}).  Together, Eqs.
\erf{genAsym} and \erf{Deltae} constitute an improved estimate of the inspiral
lifetime.

\begin{figure}
\centerline{\includegraphics[keepaspectratio=true,height=4in,angle=-90]{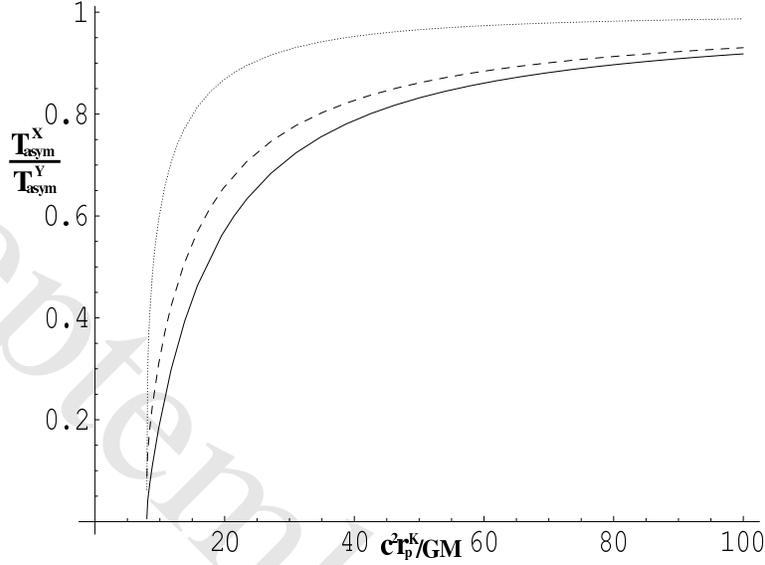}}
\caption{Comparison of the asymptotic approximation to the timescale \erf{genAsym} computed using the three methods described in the text. The plot shows the ratio $T^{Pert}_{asym}/T^{KepPM}_{asym}$ (solid line), the ratio $T^{RelPM}_{asym}/T^{KepPM}_{asym}$ (dashed line) and the ratio $T^{Pert}_{asym}/T^{RelPM}_{asym}$ (dotted line), as a function of the initial Keplerian periapse.}
\label{tAsymvsPM}
\end{figure}

Figure~\ref{tAsymvsPM} compares the asymptotic timescale \erf{genAsym} computed using three methods -- $T^{Pert}_{asym}$, generated using the perturbative result \erf{Deltae}, $T^{KepPM}_{asym}$, generated using the Peters and Mathews result \erf{tpetersAsym} evaluated for Keplerian (``K'') parameters and $T^{RelPM}_{asym}$, generated using the Peters and Mathews result \erf{tpetersAsym} evaluated for relativistic (``R'') parameters. The figure shows the ratio of these various timescales as a function of the initial periapse of the orbit. This periapse is the {\it Keplerian} periapse of the orbit, which is the quantity normally evaluated in stellar cluster simulations. The Keplerian Peters and Mathews timescale is then given directly by \erf{tpetersAsym}, while the other timescales are given by first computing the corresponding relativistic periapse and eccentricity \erf{KepRelTransform}. The curves are actually eccentricity dependent, but curves for different eccentricities are almost indistinguishable for the eccentricities of interest, $10^{-2} \lsim 1-e^R \lsim 10^{-6}$. The figure indicates that all three approximations agree for large initial periapses, but deviate as the periapse is reduced. Comparing to the standard Keplerian timescale, $T^{KepPM}_{asym}$, for $r_p^K \lsim 80GM/c^2$, the perturbative timescale is smaller by $10\% 
$ or more, while for $r_{p}^K \lsim 16GM/c^2$, it is more than $50 \% 
$ lower. This is a reasonably large discrepancy, and it therefore seems plausible that inclusion of the more accurate decay timescale in stellar cluster simulations could enhance the capture rate. On a note of caution, a parabolic Keplerian orbit with periapse $r_p^K = 8\,GM/c^2$ corresponds to the relativistic orbit with periapse $r_p^R =4\,GM/c^2$, i.e., the separatrix orbit. Thus, all orbits with Keplerian periapse less than $8\,GM/c^2$ are plunging orbits. The standard cut off used in most stellar cluster simulations is at a Keplerian periapse of $2\,GM/c^2$. This correction will thus tend to {\it decrease} the number of capture orbits. Whether this dominates over the enhanced rate due to the reduction in inspiral lifetime is not clear, but can be determined by simulation. This is currently being pursued.

It is also clear from Figure~\ref{tAsymvsPM} that a significant part of the improvement derives from the coordinate choice, i.e., using the relativistic periapse and eccentricity \erf{KepRelTransform}. A significantly improved estimate of both the radiation fluxes and the inspiral timescale can be obtained simply by evaluating the standard Peters and Mathews results (\ref{PMpardEdL}, \ref{tpetersAsym}) for the relativistically defined periapse and eccentricity \erf{KepRelTransform}. This is discussed briefly in Section~\ref{genorbits} and in more detail in \cite{GKL2}. Nonetheless, the perturbative timescale is still more than $20$\%
shorter for $r_{p}^K \lsim 16GM/c^2$, and should be used if possible. In \cite{hopman05}, relativistic parameters are used to describe the orbit and the cut off for plunging orbits is correctly defined. This might explain some of the reduction in event rate that they observe, although this reduction is dominated by diffusion onto plunging orbits. Inclusion of the perturbative results described here in the same type of simulation used in \cite{hopman05} should lead to an enhancement in rate, but it is not entirely clear how large an effect this will be.

An important point to note is that both expression \erf{genAsym} 
and the Peters expression \erf{tpetersAsym} are derived by 
integrating the orbital averaged fluxes, $\left<\rmd e/\rmd t\right>$ 
and $\left<\rmd r_{p}/\rmd t\right>$. In the test particle (zero mass) limit 
this is correct, but for non-zero mass ratios it will not be entirely 
accurate. The discrepancy is apparent from the fact that the 
gravitational decay timescale diverges like $(1-e^R)^{-1/2}$ as $e^R 
\rightarrow 1$, which is less rapid than the divergence of the 
orbital period, $(1-e^R)^{-3/2}$. If the particle was initially at 
periapse, the decay timescale is not too inaccurate, but in practice the 
particle will start near apoapse, out in the stellar cluster. 
Physically, there are no significant gravitational radiation 
losses until the particle gets close to the black hole, so the decay 
timescale must be at least as long as half the first orbital period. 
The discreteness of the GW emission should become important when one minus the initial eccentricity of the orbit, $1-e_0^R$, is less than the change in eccentricity on the first pass, $\Delta e^R(r_p^R, e^R=1)$, i.e., when
\begin{equation}
(1-e_0^R) \lsim -2\,\frac{m}{M} \,\frac{c^2\,r_p^R}{GM}\,F_{E}\left(\frac{c^2\,r_{p}^R}{GM}\right) \label{breakdown}.
\end{equation}
For this eccentricity, the change in $(1-e^R)$ over the first orbit is of the same magnitude as $(1-e_0^R)$, thus the underlying assumption that the particle completes an entire orbit on the initial geodesic is false. This is also the point at which the decay timescale \erf{genAsym} becomes comparable to the initial orbital period, so it is clear that the assumptions are breaking down. In this regime, the gravitational wave decay timescale is more accurately computed by assuming the periapse and eccentricity change discretely at periapse, and adding up the orbital periods of this sequence of geodesics. At the same level of approximation used to derive \erf{genAsym}, the corresponding GW inspiral timescale is given by
\begin{eqnarray}
\tau_{gw} (r_p^R,e_0^R) &\approx& \frac{\pi}{\sqrt{GM}} \, \left(\frac{r_p^R}{1-e_0^R} \right)^{\frac{3}{2}} + \frac{2\,\pi\,{r_p^R}^{\frac{3}{2}}}{\sqrt{GM}}  \sum_{l=1}^{\infty} \frac{1}{1-e_0^R+l\,\Delta e^R(r_p^R,e^R=1)} \nonumber \\ &=& \frac{\pi\,{r_p^R}^{\frac{3}{2}}}{\sqrt{GM}} \left(\left(\frac{1}{1-e_0^R}\right)^{\frac{3}{2}} + \left(\frac{1}{\Delta e^R(r_p^R, e^R=1)}\right)^{\frac{3}{2}} \, \zeta\left(\frac{3}{2}, \frac{1-e_0^R}{\Delta e^R(r_p^R, e^R=1)}\right) \right)
\label{tgwDiscrete}
\end{eqnarray}
where $\zeta(z,q)$ is the generalised Riemann zeta function. The first term in \erf{tgwDiscrete} is the time taken to reach periapse from apoapse on the first pass, while the second term is the summation of orbital periods over the subsequent sequence of geodesics. This expression neglects the change in periapse on each pass, the difference in $\Delta e^R$ on each pass due to the evolution of $e^R$ and approximates a finite series (which terminates when $1-e_0^R + l\,\Delta e^R$ equals the plunge eccentricity) with an infinite sum. However, these are all lower order corrections in the mass ratio, $m/M$, and may be neglected for initially highly eccentric extreme mass ratio inspirals. 

\begin{figure}
\centerline{\includegraphics[keepaspectratio=true,height=2.8in,angle=0]{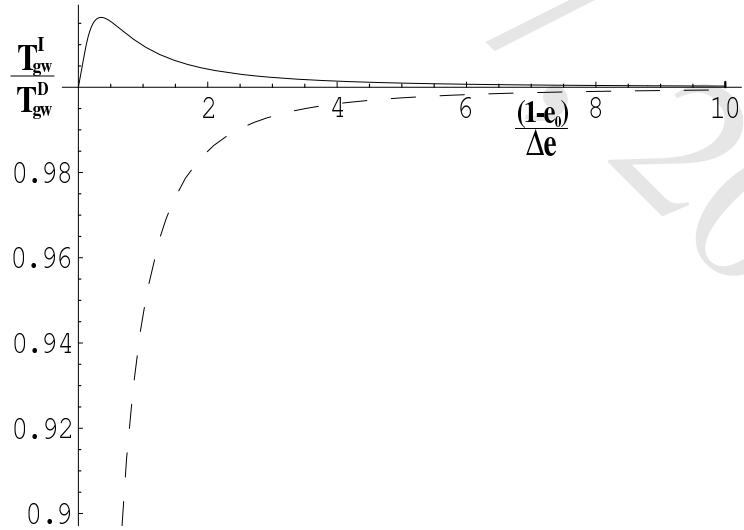}}
\caption{Ratio of the integral inspiral timescale \erf{genAsym}, $T_{gw}^{I}$, to the discrete inspiral timescale \erf{tgwDiscrete}, $T_{gw}^{D}$, as a function of $(1-e_0^R)/\Delta e(r_p^R,e^R=1)$. The dashed line uses the integral timescale computed directly from \erf{genAsym}, while the solid line includes the first half period correction \erf{corrTgw}.}
\label{DvsCfig}
\end{figure}

In the limit $1 \gg 1-e_0^R \gg \Delta e^R(r_p^R, e^R=1)$, equation \erf{tgwDiscrete} is equivalent to \erf{genAsym}, but when $1-e_0 \approx \Delta e^R(r_p^R, e^R=1)$, this is no longer true. Figure~\ref{DvsCfig} shows the ratio of the integral timescale \erf{genAsym} to the discrete timescale \erf{tgwDiscrete} as a function of $(1-e_0^R)/\Delta e^R(r_p^R,e^R=1)$. As expected, the integral form \erf{genAsym} does well until the estimated point of breakdown \erf{breakdown}, but significantly underestimates the decay timescale in the extreme parabolic limit. In this regime, a more accurate orbital decay timescale can be obtained by considering the sum of half the initial orbital period plus the decay time \erf{genAsym} evaluated for the orbit with periapse ${r_{p}^R}'=r_{p}^R+\delta r_{p}^R/2$ and eccentricity ${e^R}'=e^R+\delta e^R/2$, where $\delta r_p^R$ and $\delta e^R$ are the predicted change in periapse and eccentricity for the initial geodesic. In other words, we compute the integral decay timescale starting when the particle reaches periapse for the first time. This gives
\begin{eqnarray}
\tau_{gw} (r_{p}^R, e_{0}^R) &=& \frac{\pi}{\sqrt{GM}}\,\left(\frac{r_p^R}{1-e_{0}^R} \right)^{\frac{3}{2}} - \frac{M}{m}\,\frac{2\,\pi\,\sqrt{G\,M}}{c^2\,F_{E}\left(\frac{c^{2}\,{r_{p}^R}'}{GM}\right)} \,\sqrt{\frac{{r_{p}^R}'}{1-{e^R}'}} \nonumber \\
{e^R}' &=& e^R+ \frac{m}{M} \,\frac{c^2\, r_{p}^R}{GM}\,F_{E}\left(\frac{c^2 \,r_p^R}{GM}\right)\nonumber \\
{r_p^R}'&=&r_p^R + \frac{m}{M} \left(\frac{\sqrt{GM}(c^2r_{p}^R-2GM)^{\frac{3}{2}}}{\sqrt{2}c^2(c^2r_{p}^R-4GM)} \,F_{L_{z}}\left(\frac{c^2\,r_p^R}{GM}\right)  \right. \nonumber \\ && \left. - \frac{(r_p^R)^2\,c^2(c^2r_p^R-2GM)}{2GM(c^2r_p^R-4GM)} \,F_{E}\left(\frac{c^2 \,r_p^R}{GM}\right)\right)
\label{corrTgw}
\end{eqnarray}
Figure~\ref{DvsCfig} also shows the ratio of the revised timescale \erf{corrTgw} to the discrete timescale \erf{tgwDiscrete} (with the approximation ${r_p^R}' = r_p^R$). This expression performs very well in all regimes, and is at most a $2\%
$ overestimate near $1-e_0 = \Delta e(r_p^R,e^R=1)/4$. In most situations the difference between \erf{genAsym} and \erf{corrTgw} is small, but for any initial eccentricity, there is a mass ratio where condition \erf{breakdown} holds, and in that regime \erf{corrTgw} must be used. However, this expression is equally easy to evaluate in numerical codes.

\section{Extension to generic orbits}\label{genorbits}
This paper has focused on parabolic orbits, as these are of most relevance for astrophysical captures. For orbits of moderate eccentricity, this analysis will break down. If data based on perturbative calculations was available for the energy and angular momentum losses on generic orbits, it would be possible to compute a fit analogous to \erf{genfitform} that could be used generically (see discussion in \cite{GKL2}). This is not true at present. However, a significantly better approximation to the fluxes can be obtained simply by evaluating the Peters and Mathews fluxes, using the relativistic (rather than the Keplerian) definition of the orbital parameters. This approach was used in \cite{hopman05}.

Inversion of equations \erf{EofrpeRel}--\erf{LofrpeRel} yields a quadratic to give $e^R$ and $r_p^R$ as functions of $E$ and $L_z$ (which can be obtained from the Keplerian parameters using \erf{EofrpeRel}--\erf{LofrpeRel} if necessary). For orbits of moderate eccentricity, $e \lsim 0.9$, it is not appropriate to use the (eccentricity independent) flux expressions or timescale formula quoted earlier, since these were evaluated in the high eccentricity limit. However, if the relativistic eccentricity and periapse are computed for the orbit in question, a good approximation to the energy and angular momentum radiated can then be obtained using the Peters and Mathews expressions
\begin{eqnarray}
\rmd E &=& -\frac{64\,\pi\,c^2}{5} \, \frac{m}{M} \,
   \frac{1}{\left(1+e^R\right)^{\frac{7}{2}}}\,\left(1+\frac{73}{24}\,
   (e^R)^{2}+\frac{37}{96}\,(e^R)^{4}\right)\,
   \left(\frac{c^2\,r_{p}^R}{GM}\right)^{-\frac{7}{2}} \label{PMdE} \\
   \rmd L_{z} &=& -\frac{64\,\pi}{5}\,\frac{G\,m}{c}\,
   \frac{1}{\left(1+e^R\right)^{2}}\,
   \left(1+\frac{7}{8}\,(e^R)^{2}\right)\,\,
   \left(\frac{c^2\,r_{p}^R}{GM}\right)^{-2} \,.  \label{PMdL}
\end{eqnarray}
The corresponding inspiral timescale can be computed by integrating these fluxes along an inspiral trajectory, as in \erf{petersTgw}. We emphasise that for most situations of astrophysical interest in the capture problem, the parabolic results given earlier in this paper should be used. However, if fluxes for moderate eccentricity orbits are required, expressions (\ref{PMdE}\--\ref{PMdL}) evaluated for the relativistic orbital parameters perform much better in the strong field than if they are evaluated for the Keplerian orbital parameters \cite{GKL2}. That this approach will yield a better estimate of the energy and angular momentum fluxes is not clear a priori, but has been verified by comparing to fluxes computed using perturbation theory \cite{GKL2}. This technique essentially identifies orbits that are geometrically similar. In the strong field, a Schwarzschild geodesic with a given energy and angular momentum is nothing like the Keplerian orbit with parameters ($e^K, r_p^K$), but it does look somewhat like a Keplerian orbit with parameters ($e^R, r_p^R$), e.g., the turning points of the motion are at the same radii etc. The orbital geometry is very important for determining the radiation field and this is probably the reason that a better estimate of the flux can be obtained via this procedure.

If more accurate fluxes are required for generic orbits, one can use the expressions quoted in \cite{leor04}. These are based on post-Newtonian expansions and are therefore only approximations to the true fluxes, but improve slightly on (\ref{PMdE}--\ref{PMdL}). For parabolic orbits, the fit presented here (\ref{genfitform}\--\ref{martelfitcoeffs}) gives the correct energy and angular momentum flux from an extreme mass ratio orbit under the assumption of adiabaticity. It is therefore more accurate than any post-Newtonian calculation for parabolic and highly eccentric orbits, and so should be used under those circumstances. Further flux expressions are given in \cite{gair05}, which combine fits to perturbative calculations for the fluxes from circular orbits with post-Newtonian expressions for the fluxes from eccentric orbits. The resulting expressions improve on \cite{leor04} and could be combined with (\ref{genfitform}\--\ref{martelfitcoeffs}) to interpolate the evolution of generic inspirals.

\section{Discussion}\label{sec:Discussion}
In this paper, we have presented new simple analytic expressions for the energy and angular momentum radiated in gravitational waves by and subsequent inspiral lifetime of stars that pass close to black holes on nearly parabolic orbits. These expressions are based on the results of numerical perturbative calculations that are available in the literature \cite{martel04} and are considerably more accurate than the standard Keplerian results of \cite{PM63} that are commonly used. We find that the inspiral lifetime can be significantly reduced when the capture problem is treated more carefully, which suggests an increase in the capture rate compared to the results of current simulations. However, the use of relativistic parameters for describing the orbit might actually lead to a reduction in events, since there are more plunging orbits. 

Standard stellar cluster simulations \cite{Freitag01,Freitag02,Freitag03} characterise capture as the point at which the gravitational wave inspiral timescale becomes comparable to the two body scattering timescale. Expression \erf{corrTgw} can be easily implemented in this context in place of the usual Peters and Mathews timescale \erf{petersTgw}. More recently \cite{hopman05} used Monte Carlo simulations to study the capture problem, but allowed for diffusion by two body scattering after gravitational wave emission had become important. They found that while the standard criterion, $\tau_{gw} = (1-e)\,\tau_{rlx}$, is a reasonable approximation, there was a significant effect from scattering after this point, with many stars being perturbed onto plunge orbits rather than capture orbits. However, once $\tau_{gw} \lsim 0.01 (1-e)\,\tau_{rlx}$, this effect was unimportant. At this point, the orbits are still in general extremely eccentric, and so the parabolic flux expressions (\ref{genfitform}\--\ref{martelfitcoeffs}) and inspiral timescale \erf{genAsym} are valid and improve significantly over Peters and Mathews. Thus, the improvements presented here could also be implemented easily in this type of diffusion calculation. In the future, we hope that the results in this paper will be usefully implemented in existing stellar cluster simulation codes to investigate what effect a more careful treatment of the gravitational radiation emission can have on capture rates.

The results presented here are strictly valid only for parabolic orbits, but perform well for any orbit with sufficiently high eccentricity ($e \gsim 0.9$). We expect that in the capture problem, all orbits of interest will be highly eccentric. However, there are other astrophysically interesting scenarios in which inspiralling objects will begin on orbits with moderate or zero eccentricity. These include the formation of stars in an accretion disc around a black hole \cite{levin03,goodman03}, or the capture of stars by stripping of binaries in three body encounters \cite{miller05}. In section~\ref{genorbits} we described a simple trick that can be used to improve the accuracy of the Peters and Mathews approximation to the gravitational radiation fluxes for orbits of moderate eccentricity. Simply by using a different parameterisation of the orbit and evaluating the usual flux expression for those parameters, significantly more accurate results can be obtained \cite{GKL2}. The asymptotic approximation to the timescale \erf{genAsym} is no longer accurate when the initial eccentricity is moderate. However, capture is usually not the interesting question in astrophysical scenarios where this occurs, since the stars have already been brought onto close orbits by other mechanisms and so it will not usually be necessary to evaluate the capture criterion $\tau_{gw} < (1-e) \, \tau_{rlx}$. If an inspiral timescale is required to determine the subsequent evolution, or parameter distribution of LISA sources, this may be computed by integrating the flux expressions along the inspiral trajectory. An important caveat is that the moderate eccentricity results quoted in Section~\ref{genorbits} do not approach the parabolic results (\ref{genfitform}--\ref{martelfitcoeffs}) in the limit $e\rightarrow 1$, except in the weak field, $r_p \rightarrow \infty$. This is simply because the parabolic results are based on accurate perturbative calculations, while the results in section~\ref{genorbits} are approximations. For this reason, it would be unwise to combine both approaches in any single calculation. However, generally speaking the astrophysical situations in which the parabolic results are applicable are quite distinct from those in which the moderate eccentricity approximations are required. If it is necessary to follow an inspiral from capture right up to plunge, a scheme should interpolate appropriately between the parabolic expression (\ref{genfitform}--\ref{martelfitcoeffs}) and either the moderate eccentricity expressions quoted in Section~\ref{genorbits} or other schemes for evolving moderate eccentricity EMRIs (e.g., Barack \& Cutler 2004, Gair \& Glampedakis 2005).

The flux expressions (\ref{genfitform}--\ref{martelfitcoeffs}) scale linearly with the mass ratio, $m/M$. This follows from the assumption of an extreme mass ratio, $m/M \ll 1$. As the mass ratio is increased, the approximations used here break down in various regimes. For highly eccentric orbits, the assumption that the gravitational wave emission can be averaged over the orbit no longer holds, and it is necessary to account for the fact that the emission occurs in short bursts near periapse. This was discussed at the end of section~\ref{sec:Timescales}, and it can be accounted for in a reasonably simple way. A further consequence of increasing mass ratio is the failure of the adiabatic approximation. The energy and angular momentum fluxes are computed under the assumption that the source orbit is a geodesic of the spacetime. This is a reasonable assumption provided the timescale over which the orbital parameters change appreciably due to gravitational wave emission is long compared to the orbital timescale. This assumption breaks down if the mass ratio is too high or for orbits that lie close to the separatrix (for which the energy and angular momentum losses diverge). Roughly speaking, the adiabatic approximation breaks down when the change in eccentricity/periapse on a single encounter with the black hole is a significant fraction of the orbital eccentricity/periapse, but typically this only occurs close to plunge. The other problem at high mass ratio is the break down of the perturbative approach \--- the expressions (\ref{genfitform}--\ref{martelfitcoeffs}) are based on a fit to calculations that have been carried out to leading order in the mass ratio. As the mass ratio become moderate, this is no longer sufficiently accurate. Broadly speaking, our results should apply to mass ratios less than $\sim 10^{-2} \-- 10^{-1}$.

The results in this paper apply to orbits in the Schwarzschild spacetime, while theoretical models \cite{volon05} and some observational evidence \cite{miniutti04,fabian05} indicate that most astrophysical black holes will have significant spins. While some perturbative results are available that compute the radiation from orbits around spinning black holes \cite{poisson04}, there is not yet sufficiently generic data available from state of the art computations to fit for that situation. However, the arguments that led to the construction of the fitting function that performs so well in this case also apply when the central black hole has spin. Once a sufficient quantity of data is available, it should be possible to construct a fit of similar form, although it will be more complicated as the fit will depend on three parameters \--- the black hole spin, the radius of the periapse and the inclination of the orbit. For more generic applicability, eccentricity can also be included as a fourth parameter, though again this can only be done once perturbative calculations for generic orbits have been completed.

\acknowledgments We thank Marc Freitag for useful discussions and comments on the manuscript. SLL and JRG thank the Aspen Centre for Physics for hospitality while the manuscript was being finished. This work was supported in part by NASA grants NAG5-12834 (JRG, DJK) and NAG5-10707 (JRG) and by St.Catharine's College, Cambridge (JRG). SLL acknowledges support at Penn State from the Center for Gravitational Wave Physics, funded by the NSF under cooperative agreement PHY 01-14375, as well as support from Caltech under LISA contract number PO 1217163.

\end{document}